\documentclass{article}
\usepackage{emulateapj,pstricks}
\def\exosat     {{\em EXOSAT}\/}
\def\ginga      {{\em Ginga}\/}
\def\asca       {{\em ASCA}\/}
\def\heao       {{\em HEAO}\/-1}
\def\einstein   {{\em Einstein}\/}
\def\spacelab   {{\em Spacelab}\/-2}
\def\rosat      {{\em ROSAT}\/}
\def\am         {$^\prime$}
\def\as         {$^{\prime\prime}$}
\def\deg        {$^{\circ}$}
\def\kmsmpc     {~km$\;$s$^{-1}\,$Mpc$^{-1}$}
\def\ergs       {~erg$\;$s$^{-1}$}
\def\msun       {~$M_{\odot}$}
\def\cmsq       {~cm$^{-2}$}
\def\kms        {~km$\;$s$^{-1}$}
\def\lax	{\lesssim}

\begin{document}

\submitted{ApJ in press; 1997 July 10 revision. astro-ph/9705145}

\lefthead{DARK MATTER IN A2256}
\righthead{MARKEVITCH \& VIKHLININ}

\title{DARK MATTER AND BARYON FRACTION AT THE VIRIAL RADIUS IN ABELL 2256}

\author{Maxim Markevitch\altaffilmark{1} and Alexey Vikhlinin\altaffilmark{1}}

\affil{Harvard-Smithsonian Center for Astrophysics, 60 Garden St.,
Cambridge, MA 02138; maxim, alexey @head-cfa.harvard.edu}

\altaffiltext{1}{Also  Space Research Institute, Russian Academy of
Sciences}

\begin{abstract}

We combine high quality \asca\ and \rosat\ X-ray data to constrain the
radial dark matter distribution in the primary cluster of Abell 2256, free
from the assumption of gas isothermality. Both instruments indicate that the
temperature declines with radius. The region including the central galaxy
has a multicomponent spectrum, which results in a wide range of allowed
central gas temperatures. We find that the secondary subcluster has a
temperature and luminosity typical of a rich cluster; however, the \asca\
temperature map shows no signs of an advanced merger in this double system.
It is therefore assumed that the primary cluster is in hydrostatic
equilibrium. The data then require dark matter density profiles steeper than
$\rho\propto r^{-2.5}$ in the cluster outer part. Acceptable models have a
total mass within $r=1.5\, h^{-1}$~Mpc (approximately the virial radius) of
$6.0\pm 1.5\times 10^{14}\, h^{-1} $\msun\ at the 90\% confidence. This is
about 1.6 times smaller than the mass derived assuming isothermality. The
gas fraction is correspondingly higher and is $0.08\pm 0.02\, h^{-3/2}$. A
lower limit on the fraction of gas in the total local density at the same
radius is $0.09\,h^{-3/2}$, which is twice the isothermal value. Near the
center, dark matter profiles with and without central cusps are consistent
with the data. Our inferred total mass inside the X-ray core
($r=0.26\,h^{-1}$ Mpc) is $1.28\pm 0.08\times 10^{14}\, h^{-1}$\msun, which
exceeds the isothermal value by a factor of 1.4. Although the confidence
intervals above may be underestimates since they do not include
uncertainties arising from asymmetry and departures from hydrostatic
equilibrium, the behavior of the mass distribution, if applicable to other
clusters, can bring into better agreement X-ray and lensing mass estimates,
but aggravate the ``baryon catastrophe''. The observed considerable increase
in the gas content with radius, not anticipated by simulations, may imply
that a significant fraction of thermal gas energy comes from sources other
than gravity and merger shocks, such as supernovae-driven galactic winds,
for example.

\end{abstract}


\section{INTRODUCTION}

It has long been suggested that measuring the density and temperature
distributions of the intergalactic gas would allow a cluster mass
determination free from the limitations inherent in the virial estimates
based on galactic velocities (Bahcall \& Sarazin 1977; Mathews 1978). The
necessary requirement that the gas be in hydrostatic equilibrium in the
cluster gravitational well is largely met in most regular clusters (e.g.,
Sarazin 1988; Navarro, Frenk, \& White 1995; this issue is discussed in more
detail in \S4.2). The gas density profile for symmetric clusters can readily
be obtained with an imaging instrument, such as \einstein\ or \rosat. Obtaining
temperature distributions has proven to be more problematic. In the absence
of spatially resolved spectral data, much work has been done assuming
constant temperature. At small cluster radii and outside of the central
cooling flow regions, this assumption indeed applies (e.g., Watt et al.\
1992; Ikebe et al.\ 1996; Loewenstein \& Mushotzky 1996a). However, mass
estimates at large radii are only as good as the temperature measurements at
those radii.  This is illustrated by large uncertainties of the inferred
mass when allowance for nonisothermality is made (e.g., Henry, Briel, \&
Nulsen 1993, hereafter HBN, obtain a 68\% range spanning a factor of three
for the A2256 mass within $r_{\rm vir}$; see also Loewenstein 1994).
Cosmological simulations predict that cluster gas temperatures decline at
large radii (e.g., Tsai, Katz, \& Bertschinger 1994; Navarro et al.\ 1995;
Evrard, Metzler, \& Navarro 1996).  At the time of this writing, however,
only Coma and A2163 have temperature measurements at radial distances
exceeding $1\,h^{-1}$ Mpc.

The \asca\ X-ray satellite (Tanaka, Inoue, \& Holt 1994) is capable of
measuring the cluster temperature spatial distributions. It indeed observes
a significant temperature decline at large off-center distances in a number
of hot, massive systems, such as A2163 (Markevitch et al.\ 1996, hereafter
M96a), A2256, A2319, A665 (Markevitch 1996, hereafter M96); A2218
(Loewenstein 1997); and probably, in the cooler A3558 (Markevitch \&
Vikhlinin 1997, hereafter MV) and Hydra-A (Ikebe et al.\ 1997). Earlier
\exosat\ and \ginga\ data on Coma (Hughes, Gorenstein, \& Fabricant 1988;
Hughes 1991) and \spacelab\ results on Perseus (Eyles et al.\ 1991)
suggested qualitatively similar temperature behavior. Henriksen \& White
(1996) infer a radial temperature decline for several clusters from
comparison of spectra obtained with \heao\ and \einstein\ SSS with different
fields of view. If declining temperature profiles are typical (as is indeed
suggested by our \asca\ work in progress), it would imply that there is in
fact less dark matter in cluster outskirts and probably more in the inner
parts compared to the results derived under the isothermal assumption. In
this paper, we will quantify this effect for A2256 and discuss some of its
implications.

\begin{figure*}[bh]
\small
\renewcommand{\arraystretch}{1.3}
\renewcommand{\tabcolsep}{6mm}
\begin{center}
\vspace{-3mm}
TABLE 1
\vspace{2mm}

{\sc Parameters of image components from \rosat\ PSPC}
\vspace{2mm}

\begin{tabular}{lccccc}
\hline \hline
{\sc Component} & \multicolumn{2}{c}{\sc Centroid position (J2000)} &
$L_x\,h^2$ (0.5--2 keV), & \multicolumn{2}{c}{\sc Density profile} \\ 
          & R.A. & Dec. & \ergs  &  $a_x$ & $\beta$ \\
\hline
Primary cluster & $17^{\rm h} 04^{\rm m} 35^{\rm s}\!.8$ & 78\deg 36\am 54\as &
$8.9\times 10^{43}$ & $5'\!.64$ & 0.816 \\
Subcluster      & $17^{\rm h} 03^{\rm m} 11^{\rm s}\!.0$ & 78\deg 39\am 40\as &
$4.3\times 10^{43}$ & $3'\!.9$ & 0.92 \\
Central source$^1$ & $17^{\rm h} 04^{\rm m} 29^{\rm s}\!.0$ & 78\deg 38\am
35\as & $0.45\times 10^{43}$ & ... & ... \\
\hline
\end{tabular}
\vspace{2mm}

\begin{minipage}{15.5cm}
$^1$ Coordinates are fixed at the optical position of the brightest galaxy;
Gaussian brightness profile width is fixed at $\sigma=1'$.
\end{minipage}
\end{center}
\end{figure*}

A2256 ($z=0.058$) is among those clusters with a detected temperature
decline. Although it may not be the best candidate for a convincing mass
measurement due to the presence of substructure (discussed in \S4.1), the
quality of the available X-ray data for this cluster is unsurpassed.  There
are six deep \rosat\ PSPC pointings covering different cluster regions out
to $r\simeq 3\,h^{-1}$~Mpc (Briel et al.\ 1991, hereafter B91) and a high
statistical quality \asca\ temperature map for $r<1\,h^{-1}$~Mpc with useful
spatial resolution (M96).  A2256 is therefore a good starting point for a
mass determination using a measured temperature profile. A similar earlier
effort on the more distant A2163 (M96a) has suggested that the dark matter
density in that cluster falls off rather steeply with radius --- but also
suggested that A2163 may be out of hydrostatic equilibrium at the radii of
interest. More evidence has since been added by Squires et al.\ (1997) that
A2163 indeed is an advanced merger. Below, essentially the same method as
that used in M96a is applied to A2256, which is better resolved and more
likely to be in hydrostatic equilibrium. We parameterize $H_0\equiv
100\,h$\kmsmpc; all quoted errors are 90\% one-parameter intervals.

\section{\rosat\ DATA}

B91 have shown that A2256 is a double X-ray cluster. In addition, one of the
X-ray brightness peaks coincides with the brightest cluster galaxy and is
displaced by about 1.7\am\ from the primary cluster's centroid. To derive
the density profile of the primary cluster, we have fitted the \rosat\ PSPC
image by two symmetric projected $\beta$-models (e.g., Jones \& Forman 1984)
representing the two subclusters, their $a_x$, $\beta$, center positions and
relative normalizations and the X-ray background being free parameters, and
a Gaussian brightness peak at the position of the central galaxy (fixing
$\sigma=1'$ as suggested by the HRI image). \rosat\ PSPC data are analyzed
as described in MV. For image analysis, only energies 0.7--2 keV were used
to optimize statistics, and the \rosat\ PSF model was taken from Hasinger et
al.\ (1993). The best-fit centroid coordinates, luminosities and gas
distribution parameters for the components of the fit are summarized in
Table 1.

For the primary cluster, we obtain $a_x=5'\!.64=0.259\,h^{-1}$ Mpc and
$\beta=0.816$ for the emission measure profile (confidence contours are
shown in Fig.\ 1), taking into account the small effect of the observed
temperature decline (M96 and \S3) and excluding the detectable point
sources. These parameters are derived using the data from $r<25'$ for
self-consistent use in the \asca\ analysis below. The corresponding central
electron density (excluding the central galaxy) is $n_{e0}=3.1 \pm 0.1
\times 10^{-3}\,h^{1/2}$ cm$^{-3}$. Our simplifying assumption that the
central galaxy is projected rather than embedded in the primary cluster,
introduces a small inaccuracy of the central density distribution of the
main cluster. However, this inaccuracy is insignificant comparing to the
uncertainty in the central temperature which will be obtained below; we
discuss this issue further in \S5.1.

Regarding the adequacy of the $\beta$-model at large radii, the primary
cluster emission can still be traced at the $1\sigma$ level, including the
background uncertainty, out to $r=60'$, which is twice the virial radius
(the radius inside which the cluster mean density is 180 times the closure
density; $r_{\rm vir}\simeq 1.5\,h^{-1}$ Mpc for A2256 using our mass
measurement).  Using data within $r=60'$, we obtain $a_x=5'\!.51$ and
$\beta=0.801$, not significantly different from the above values. This is
contrary to the simulation predictions of steepening gas profiles with
radius (e.g., Navarro et al.\ 1995; see, however, Frenk et al.\ 1996).
Fixing $a_x$ and using only the profile in the several smaller intervals of
radii, we find that the global value of $\beta$ represents the local density
slope to about $\pm 5$\% accuracy in the inner cluster part. At the radii
where the background uncertainty dominates, the possible error is greater:
for $r=20'-40'$, we obtain $\beta=0.83\pm 0.17$. These uncertainties will be
included in the mass estimates discussed below.

For the smaller subcluster, $a_x= 3'\!.9 \pm 0'\!.4$ and $\beta=0.92\pm
0.12$.  These parameters are poorly constrained but are not important for
our final results, thus in the following analysis they are fixed at their
best-fit values. The two subclusters' 0.5--2 keV luminosities within
$r=1\,h^{-1}$ Mpc are given in Table 1. The smaller subcluster is two times
less luminous than the primary cluster; this differs from the finding of B91
(1/5 of the primary cluster's luminosity). The difference is due to the fact
that B91 refer to an excess flux in the 90\deg\ image sector containing the
subcluster, while we refer to the full flux in the subcluster's
$\beta$-model. Fitting the PSPC spectrum of the subcluster while subtracting
the emission of the primary component similarly to B91, fixing the value of
$N_H=5\times 10^{20}$\cmsq\ and using the 0.5--2 keV band (as in MV), we
obtain $T_e=4.1$ (2.5--8.2) keV, consistently with the \asca\ result given
below.

The PSPC-measured profile of projected temperatures in the inner $r \simeq
1\,h^{-1}$ Mpc of A2256 agrees remarkably well with the \asca\ radial
temperature gradient (MV), but due to its relatively large errors is not
used in the mass analysis below. Fig.~2 shows PSPC temperature measurements
at still greater radii, $r \simeq 1-1.6\,h^{-1}$ Mpc (excluding the 90\deg\
sector with the smaller subcluster), obtained using the same procedure as in
MV.  Although poorly constrained (mostly due to the background uncertainty),
they suggest that the gas temperature continues to fall with radius. While
keeping this in mind, we will not use these data below. More important for
our analysis is the fact that the gas emission is still detected with 90\%
confidence at $r=40'\simeq 1.8\,h^{-1}$ Mpc. We will conservatively
interpret this as a lower limit on the gas temperature, $T_e>0$ at that
large radius, and apply this limit to the hydrostatic temperature profiles
in \S5.

\begin{figure*}[ht]
\pspicture(0,8.0)(18.8,20)

\rput[tl]{0}(-0.2,20.1){\epsfysize=8.9cm
\epsffile{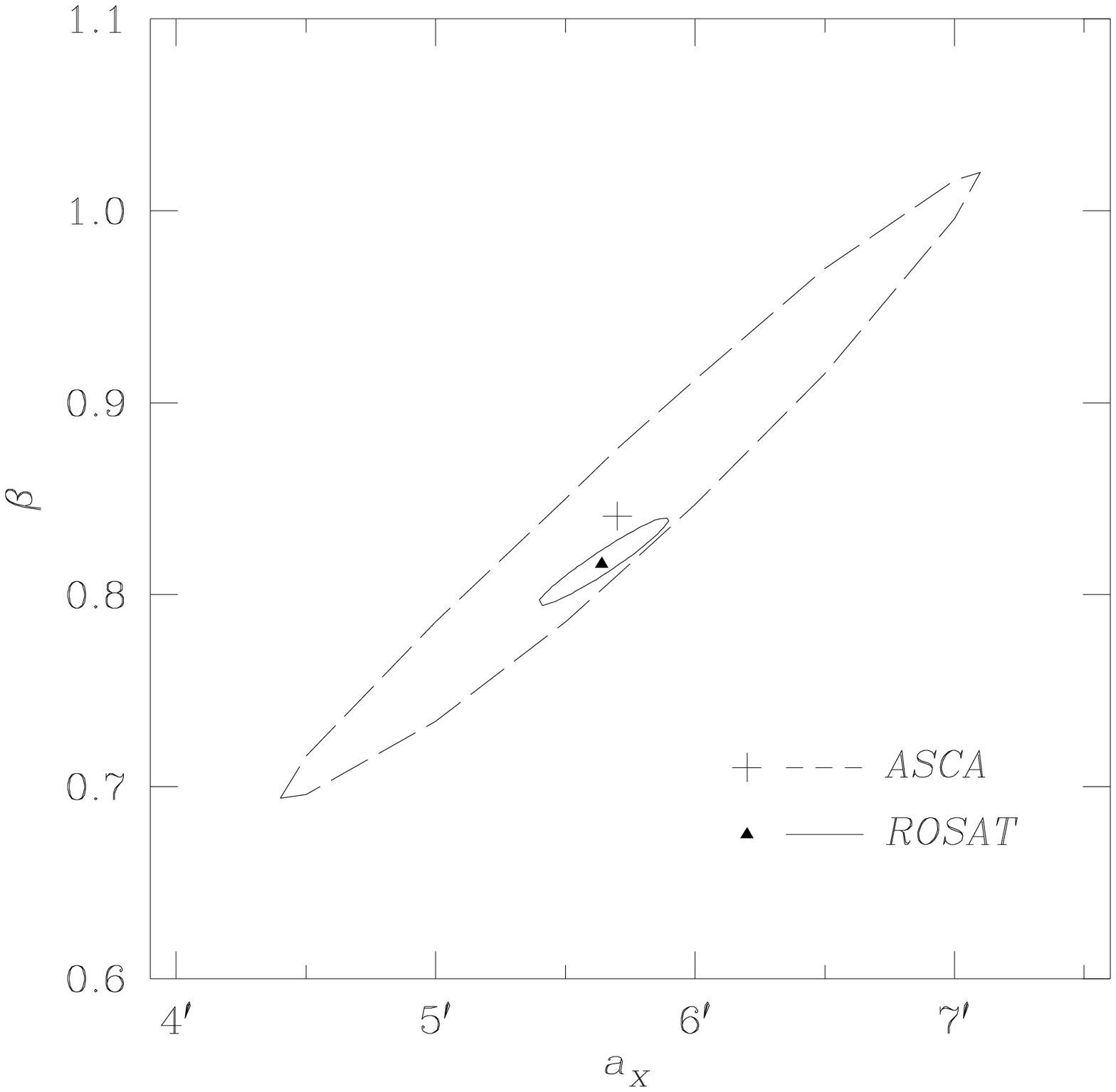}}

\rput[tl]{0}(9.5,20.1){\epsfysize=8.9cm
\epsffile{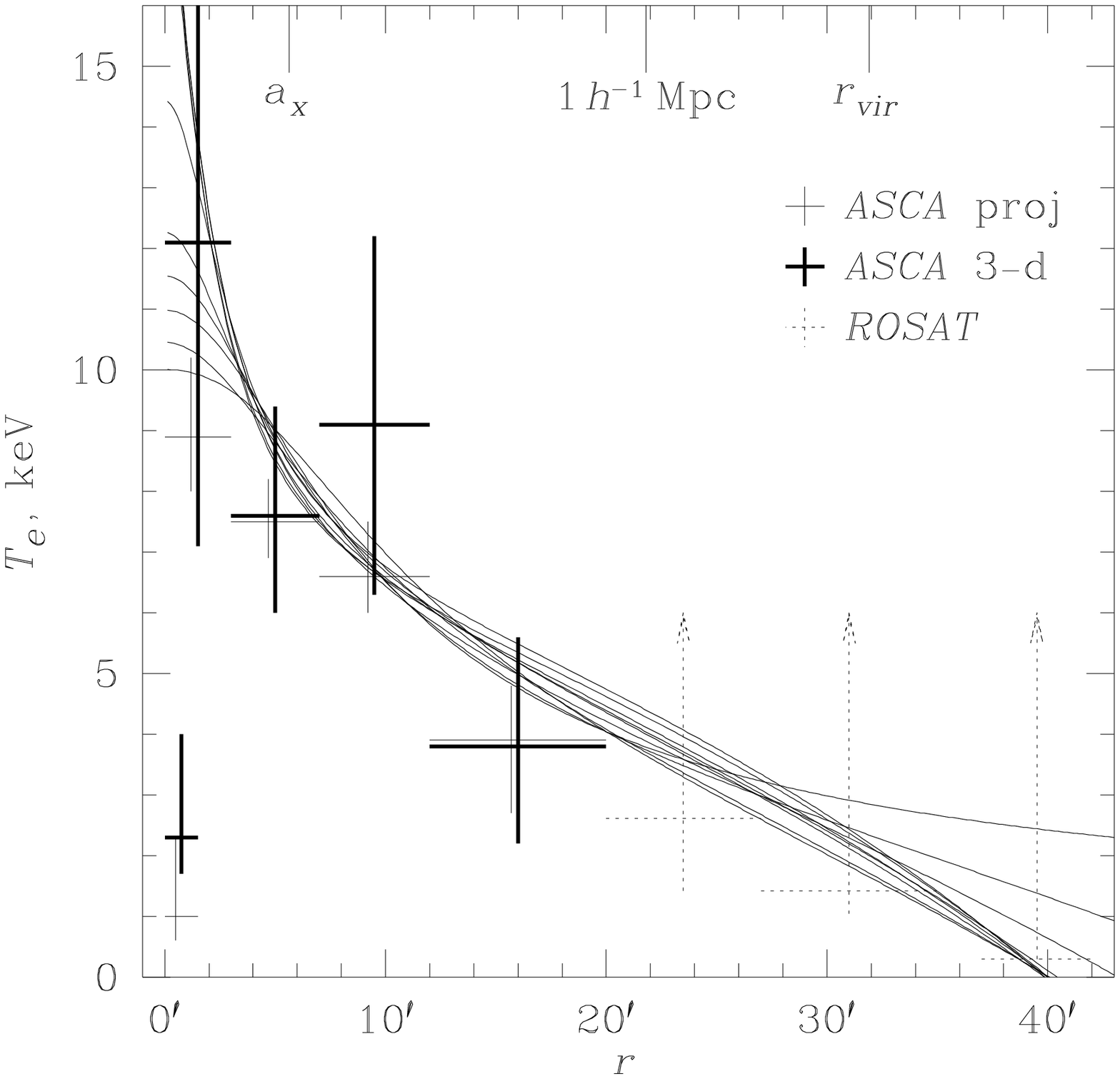}}

\rput[tl]{0}(-0.1,11.2){
\begin{minipage}{8.9cm}
\small\parindent=3.5mm
{\sc Fig.}~1.---Two-parameter 90\% confidence regions on $a_x$ and $\beta$
of the primary cluster gas density profile from \asca\ and \rosat. For
\asca, all temperatures and the spectrum of the central source were free
parameters. For both instruments, parameters $a_x$ and $\beta$ of the
secondary cluster were fixed at their best-fit values; \rosat\ fit excluded
the region of the central source.
\end{minipage}
}

\rput[tl]{0}(9.6,11.2){
\begin{minipage}{8.9cm}
\small\parindent=3.5mm
{\sc Fig.}~2.---Temperature profile of the primary cluster. \asca\ values
correspond to step-like profiles in projected annuli or spherical shells
(upper axis gives conversion to linear distances), while \rosat\ values are
for projected annuli. The central low temperature shows an additional
component required in the central region. Errors are 90\%. The lower limit
at $r=40'$ reflects the \rosat\ detection of the cluster emission there.
Smooth lines show a representative set of models (not projected and not
showing the cooler component) allowed at the 99\% confidence by the \asca\
data and the \rosat\ 40\am\ limit (see text).
\end{minipage}
}
\endpspicture
\end{figure*}

\section{\asca\ DATA}

%
%

Technical details of the \asca\ analysis of A2256, including discussion of
the systematics, are given in M96, who presented a two-dimensional map of
projected gas temperatures in this cluster. The analysis technique used
there and in this work, which efficiently takes into account the complex
\asca\ PSF (Takahashi et al.\ 1995), is described in M96a. The cluster
temperature map shows that except for a cooler subcluster, there is no
significant asymmetry in the radial decline of the projected temperature.
Here we reconstruct a three-dimensional, as well as projected, radial
temperature profile of the primary cluster. In this Section, we make no
assumptions regarding the underlying mass distribution to indicate the
intrinsic data constraints, while in \S5, continuous hydrostatic profiles
will be constrained by the data. As in \S2, we assume that the main cluster
is spherically symmetric and that the smaller subcluster and the central
source are projected onto it. It is also assumed for simplicity that the
subcluster is isothermal, since it only contributes about 20\% of the flux
in the image region that determines the temperature in our most important
outer spherical bin (for the same reason, the uncertainty in its brightness
distribution is unimportant).  Under these assumptions, we first fit uniform
temperatures in four spherical shells of the primary cluster with radii
corresponding at the cluster's distance to $r=0'-3'$, $3'-7'$, $7'-12'$ and
$12'-20'$. Together with them, we also fit a temperature of the subcluster
and various models for the central source, allowing its normalization to be
a free parameter.  The image regions from which \asca\ spectra are collected
are annuli defined by the above radii excluding the 90\deg\ sector
containing most of the subcluster, this sector, and an $r=1.5'$ circle
around the central galaxy.  GIS and SIS spectra from all regions are fitted
simultaneously.

Due to the limited \asca\ angular resolution, it is difficult to constrain
both the temperature and spatial brightness distributions with \asca\ alone.
We therefore take advantage of the \rosat\ information on the relative
brightness distribution. Because the resulting \asca\ temperatures are
sensitive to the assumed brightness distribution, one must ensure
consistency between the two instruments on this issue and that the use of
the \rosat\ image does not introduce an error in the temperature
measurement. For this, we fix $a_x$ and $\beta$ of the subcluster and the
positions of the three model components at their \rosat\ values, but free
the temperatures, relative normalizations of the model components (that is,
the main cluster, the subcluster, and the central source), and $a_x$ and
$\beta$ of the primary cluster, thus using almost no \rosat\ information in
the \asca\ fit.  Confidence contours for the latter two parameters are shown
in Fig.\ 1. \rosat\ and \asca-derived brightness distributions are in
perfect agreement, while the \rosat\ accuracy is superior. Therefore, $a_x$
and $\beta$ can be fixed at their \rosat\ values.

Assuming that the central source has the same temperature as the surrounding
gas yields a gas temperature of $7.2\pm1.5$ keV in the central $r=3'$
spherical bin.  This may misleadingly suggest that the cluster central part
is largely isothermal. However, if that source is allowed to have a
different spectrum, $\chi^2$ is significantly reduced by 13--18 for one
additional parameter depending on the assumed model, for which we tried a
second thermal component, a cooling flow or a power law. A power law model
(with the best-fit photon index $-2.4\pm 0.3$) is formally the best fit, and
we chose to use this model in further analysis (being aware that the real
spectrum is probably still more complex). All models result in allowing a
higher ambient gas temperature. In fact, due to the limited \asca\ angular
resolution, we cannot precisely localize the source of this additional
emission: it may, for example, be one or more of the several radio sources
in the central region (R\"ottgering et al.\ 1994). There are no bright point
sources in the relevant region of the \rosat\ HRI image, which argues in
favor of either thermal or nonthermal extended component. However, its exact
origin is not important for our analysis, therefore we assume it is the
central galaxy. This spectral complexity, combined with projection, results
in a relatively large uncertainty in the central temperature of the main gas
component, which will complicate an accurate measurement of the mass
distribution in the central region.

The best-fit temperatures in the radial shells of the primary cluster with
other parameters treated as described above, are shown in Fig.\ 2 as thick
crosses. For this model, $\chi^2_{\rm min}=461$ for 685 d.o.f., reflecting
our conservatism in accounting for the instrumental uncertainties. For a
consistency check and to show the intrinsic data constraints separated from
the uncertainty of the deprojection, we performed a similar analysis
assuming uniform temperatures within projected regions on the sky rather
than in spherical shells (even though it has less physical meaning). These
temperatures are also shown in Fig.\ 2 as thin solid crosses. The projected
and three-dimensional temperatures differ as expected, with uncertainties of
the latter being greater due to the deprojection. Note that the step-like
profiles assumed above have little physical meaning (since the real
distribution is continuous), which, together with the natural
anticorrelation of the errors in the adjacent spherical bins of the
deprojected profile, explains differences such as that in the third annulus
in Fig.\ 2. This is not important for our mass analysis in \S5.1, in which
continuous temperature profiles will be compared directly to the data.

For the smaller subcluster we obtain $T_e=5.3\pm 0.5$ keV and a relative
iron abundance $0.35^{+0.16}_{-0.11}$, exceeding with marginal significance
the value for the primary cluster, $0.23^{+0.09}_{-0.07}$. Note that the
temperature and the luminosity of the subcluster agree very well with the
$L_x-T$ relation of David et al.\ (1993). The iron abundance in the
outermost parts of the primary cluster is unconstrained; we assume it
uniform and fix at the best-fit value.

\section{VALIDITY OF HYDROSTATIC EQUILIBRIUM ANALYSIS}

\subsection{Substructure and possible merger}

Before using X-ray data to derive the distribution of dark matter in A2256
under the assumption of hydrostatic equilibrium, we should address the
adequacy of this assumption. Fabricant et al.\ (1989), B91, and Davis \&
Mushotzky (1993) showed that A2256 contains substructure both in the optical
and X-ray data. This cluster has a radio halo and head-tail radio sources
(R\"ottgering et al.\ 1994) which may be related to a merger (although their
relation is not unambiguously established). A merger would result in
significant gas turbulence and inadequacy of our assumption of spherical
symmetry. However, there is a strong argument against an advanced merger,
which is the absence of any significant asymmetries of the gas temperature
(M96) expected in a collision of such massive clusters (e.g., Schindler \&
M\"uller 1993). The detection of hot spots by Briel \& Henry (1994) was
later not confirmed by reanalysis of their data using an updated \rosat\
PSPC calibration (MV). The accuracy of the \asca\ measurement in M96 was
adequate for detecting a temperature pattern characteristic of a merger, and
indeed such a pattern was observed in several clusters, including the
obviously merging A754 (Henriksen \& Markevitch 1996).  The observed
probable difference in iron abundances of the subclusters (\S3) also argues
against a late-stage merger since the gas would mix if clusters had passed
through each other.  Another argument is the presence of a remarkably long
and straight narrow tail radio source near the cluster center (R\"ottgering
et al.\ 1994). Its existence strongly suggests that there is no significant
gas turbulence there. We therefore assume that the merger has not proceeded
far enough to disturb the bulk of the primary cluster's gas. Moreover, below
it will be assumed for simplicity that the two clusters are physically well
separated along the line of sight, so that the gravitational effect of the
subcluster on the primary cluster's gas can be neglected. Indeed, such a
separation is suggested by the galaxy velocity data.  Fabricant et al.\
(1989) and B91 noted that the cluster line-of-sight velocity distribution
can be described by two components, correspondent to the two subunits,
separated by about 2000\kms. A simple calculation (as in, e.g., Henriksen \&
Jones 1996) shows that, assuming two clusters of comparable masses falling
from rest at some large distance toward each other, a mass of the main
cluster of $7\times 10^{14}\,h^{-1}$\msun\ as obtained below, and the
observed projected distance between the cluster centers, this rather small
relative velocity implies a separation of $2-3\,h^{-1}$ Mpc.

\subsection{More general issues}

In a similar analysis of A2163, M96a raised the possibility that plasma in
the outer, low density region may be out of electron-ion temperature
equipartition, in such a way that the measured $T_e$ underestimates the
thermodynamic temperature. In A2256, the temperature measurements correspond
to smaller radii and higher densities. In the radial bin where \asca\
detects a temperature drop, the timescale for reaching equipartition via
collisions is $\sim 7\times 10^8\,h^{-1/2}$ yr, shorter than the time which
has to have passed since the last merger for the cluster to become relaxed
as we observe it. Actual temperature measurements beyond this radius will
not be used (only a lower limit, $T_e>0$). The temperature decline in A2256
is also less steep compared to that in A2163 where it implied convective
instability. At the radii of the temperature decline in A2256, the
temperatures correspond to a polytropic index of 1.50 (1.24--1.9).
Therefore, the gas should be convectively stable on large scales.

There is a possibility that some of the gas in the outer cluster is confined
by the local potential wells of small infalling subunits, rather than that
of the primary cluster. If that is the case one would not expect the cluster
to be azimuthally symmetric, while the \rosat\ image shows it is rather
smooth out to large radii, except for a small ellipticity.

At large cluster radii, residual bulk motions of gas from past merger events
and infall may become non-negligible, which would mean that hydrostatic
equilibrium may not be an adequate assumption. Simulations of Evrard et al.\
(1996) showed that the kinetic energy of these motions is between 0--70\% of
the gas thermal energy at the cluster virial radius, with high values
corresponding to clusters experiencing mergers. Given that the bulk of A2256
seems to have been undisturbed in the recent past as we concluded above, we
can assume that A2256 is relatively well virialized within our radius of
interest. A significance of this effect, as well as the uncertainty arising
from our spherical symmetry assumption, can best be assessed through the
hydrodynamic simulations. Several independent authors (e.g., Tsai et al.\
1994; Schindler 1996; Evrard et al.\ 1996; Roettiger et al.\ 1996) applied
the X-ray mass estimation algorithm, which assumes spherical symmetry and
hydrostatic equilibrium, to the simulated clusters in various cosmological
backgrounds. They find that, when the actual temperature profile is used (as
we do), such estimates are on average unbiased and have an {\em rms} scatter
around the true mass of about 15\% at the core and about 30\% at the virial
radius. These are conservative estimates of the scatter because clusters
with obvious mergers, which show the biggest mass errors, are included in
them but would be excluded by the observers (recall that the primary cluster
of A2256 is not a merger, as we concluded above). Bartelmann \& Steinmetz
(1996) find their simulated X-ray mass estimates to be systematically
biased, and attribute the discrepancy with other authors to their assumption
of more severe observational conditions, which are not relevant to our
present work.  The values above may therefore be regarded as conservative
estimates of the systematic uncertainty of an individual cluster mass
determined by the X-ray method. We will continue this discussion in \S6.1 in
application to our results.

\section{MASS CALCULATION}

\subsection{Method}

The X-ray data will now be used to constrain the mass profile of the primary
cluster. Because the gas temperature profile is known much less accurately
than the gas density profile, we chose to use the fitting method of Hughes
(1989) and HBN with some technical variations. Two functional forms for the
dark matter radial profile will be considered, which together can
approximate just about any physically meaningful symmetric distribution: one
with a constant core,
\begin{equation}
\rho_d \propto \left( 1+\frac{r^2}{a_d^2} \right)^{-\alpha/2},
\end{equation}
and one with a central singularity (cusp) predicted by cluster simulations
(e.g., Navarro et al.\ 1995),
\begin{equation}
\rho_d \propto \left( \frac{r}{a_d} \right)^{-\eta} \left(
1+\frac{r}{a_d} \right)^{\eta-\alpha}.
\end{equation}
In both profiles, the density in the outer part declines asymptotically as
$\rho_d \propto r^{-\alpha}$. Navarro, Frenk, \& White (1997) found that
simulations in different cosmologies produce a ``universal'' density profile
for clusters in equilibrium; it is of the latter form with $\eta=1$ and
$\alpha=3$. The isothermal sphere model corresponds to $\alpha=2$; profiles
with $\alpha=2-4$ are expected under different assumptions about the halo
formation (e.g., Syer \& White 1997 and references therein; Kofman et al.\
1996). For the gas, we use a $\beta$-model density profile (Cavaliere \&
Fusco-Femiano 1976),
\begin{equation}
\rho_g= \rho_{g0} \left(1+\frac{r^2}{a_x^2} \right)^{-3\beta/2}
\end{equation}
with $a_x$ and $\beta$ derived in \S2. For simplicity, the relatively small
(of the order of 1\%, HBN) contribution of galaxies to the total mass is
neglected.  With the above functional forms for the major mass components,
we solve the hydrostatic equilibrium equation (e.g., Sarazin 1988)
analytically (see Appendix) to recover the radial gas temperature profile
for a given set of parameters $\alpha$, $\eta$ (if applicable), $a_d$, the
dark matter normalization $\rho_{d1}$, and the temperature at zero radius
$T_0$. The general behavior of the solutions is discussed in detail by e.g.,
Hughes (1989), HBN, and Loewenstein (1994). For dark matter profiles of the
cusp form (2) and a $\beta$-model gas density profile, the gas temperature
profile at $r=0$ is finite if $\eta<2$ and has zero derivative if $\eta<1$.
In practice, the insufficient spatial resolution of the temperature data
does not allow us to independently constrain all three shape parameters of
the profile (2), thus we chose to fix $\eta=1$ as suggested by simulations.
Temperature profiles rising to infinity outside the measurable region are
not allowed in our analysis (which in practice does not affect our
constraints), but we do allow them to fall to zero at a finite radius
(outside $r=40'$ where \rosat\ certainly detects cluster emission), since
hydrostatic equilibrium is not expected to hold far from the virial radius.

In this work, we are interested in the shape of the dark matter density
distribution. Therefore, for a given set of the shape parameters $\alpha$
and $a_d$, we fit $\rho_{d1}$ and $T_0$ so that after the projection and
convolution with the telescope responses, the temperature profile minimizes
\asca's $\chi^2$. The spectra and the normalizations of the additional
central galaxy source and the subcluster are treated as free parameters to
allow maximum freedom for the temperature profile of the main cluster. The
gas associated with the central galaxy is not included in the hydrostatic
equilibrium equation, although in reality, the central galaxy atmosphere
resides in the same gravitational potential and is in pressure equilibrium
with the main cluster gas that is being modeled. In principle, its
temperature can be modeled in a similar way and used to provide additional
constraints on the mass distribution in the center. However, in the absence
of better-resolution data on whether the two gas phases cohabit in the same
volume or one of them replaces the other, we chose a conservative treatment
that does not additionally restrict the range of mass models and consists of
completely freeing the normalization and the spectrum of the central galaxy
component. This approach should not significantly affect the mass
measurements outside the $r\simeq 2'$ central galaxy region.

The model is fit directly to the \asca\ data (not to the step-like
temperature profiles shown in Fig.\ 2). Minimum $\chi^2$ values for both
functional forms of the dark matter profile are very similar. They are also
similar to the value obtained for the step-like temperature distribution,
thus this value can be considered an ``absolute'' minimum of $\chi^2$ with
which to compare the models of different form. Those models for which
$\chi^2>\chi^2_{\rm min}+4.6$ are rejected with 90\% confidence.  A profile
acceptable for \asca\ is then checked against the requirement that the
temperature be greater than zero at $r=40'$ (a lower limit from the \rosat\
flux detection) and if not complying, rejected at the 90\% confidence. The
actual \rosat\ temperature estimates between $r=20'-35'$ do not constrain
profiles any better than this lower limit, thus for simplicity they are not
used. However, these estimates reinforce the lower limit even if, contrary
to our assumption, hydrostatic equilibrium does not hold at these large
radii.

\begin{figure*}[ht]
\pspicture(0,1.2)(18.8,20)

\rput[tl]{0}(-0.2,20.1){\epsfysize=8.4cm
\epsffile{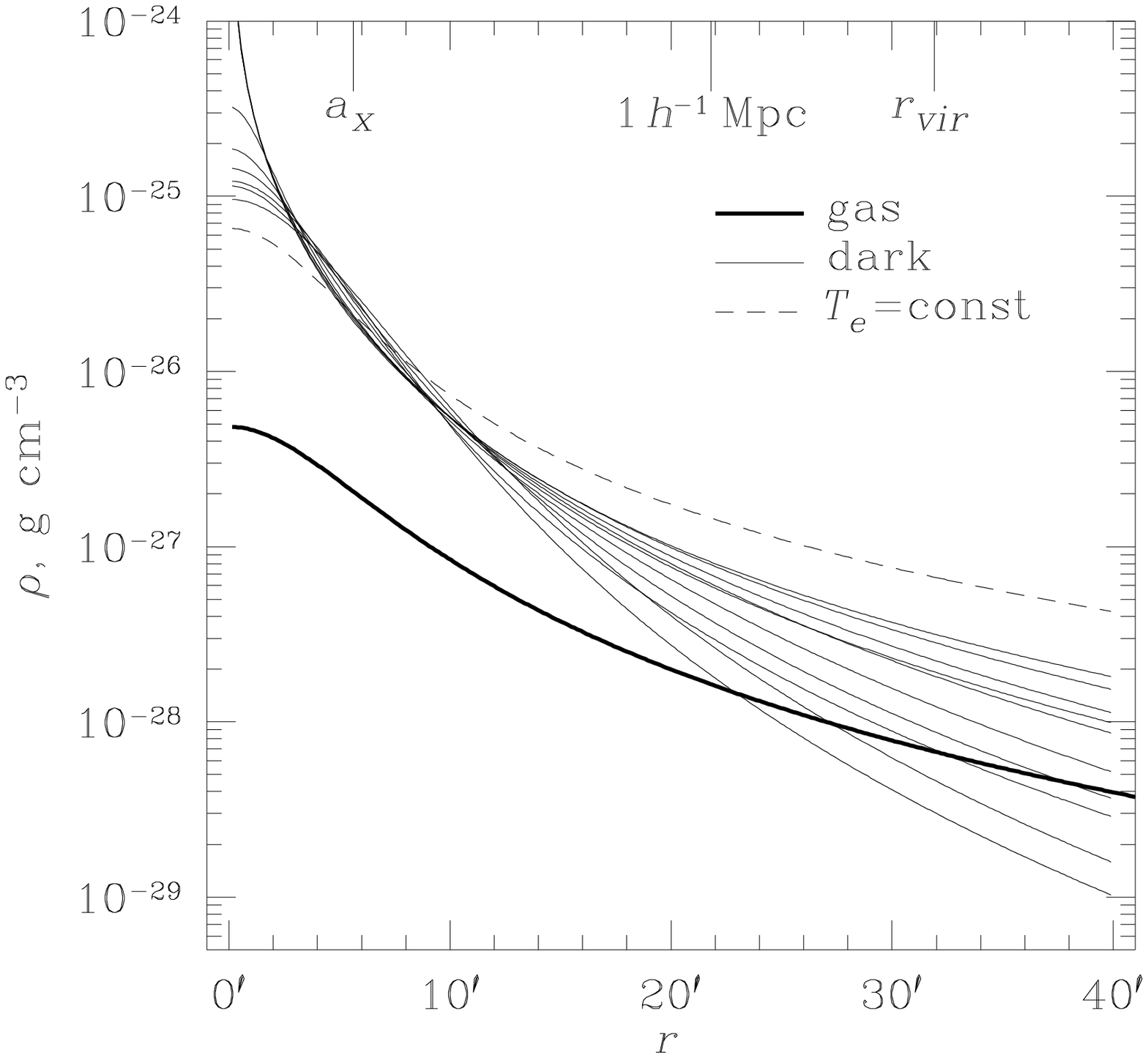}}

\rput[tl]{0}(9.1,20.1){\epsfysize=8.4cm
\epsffile{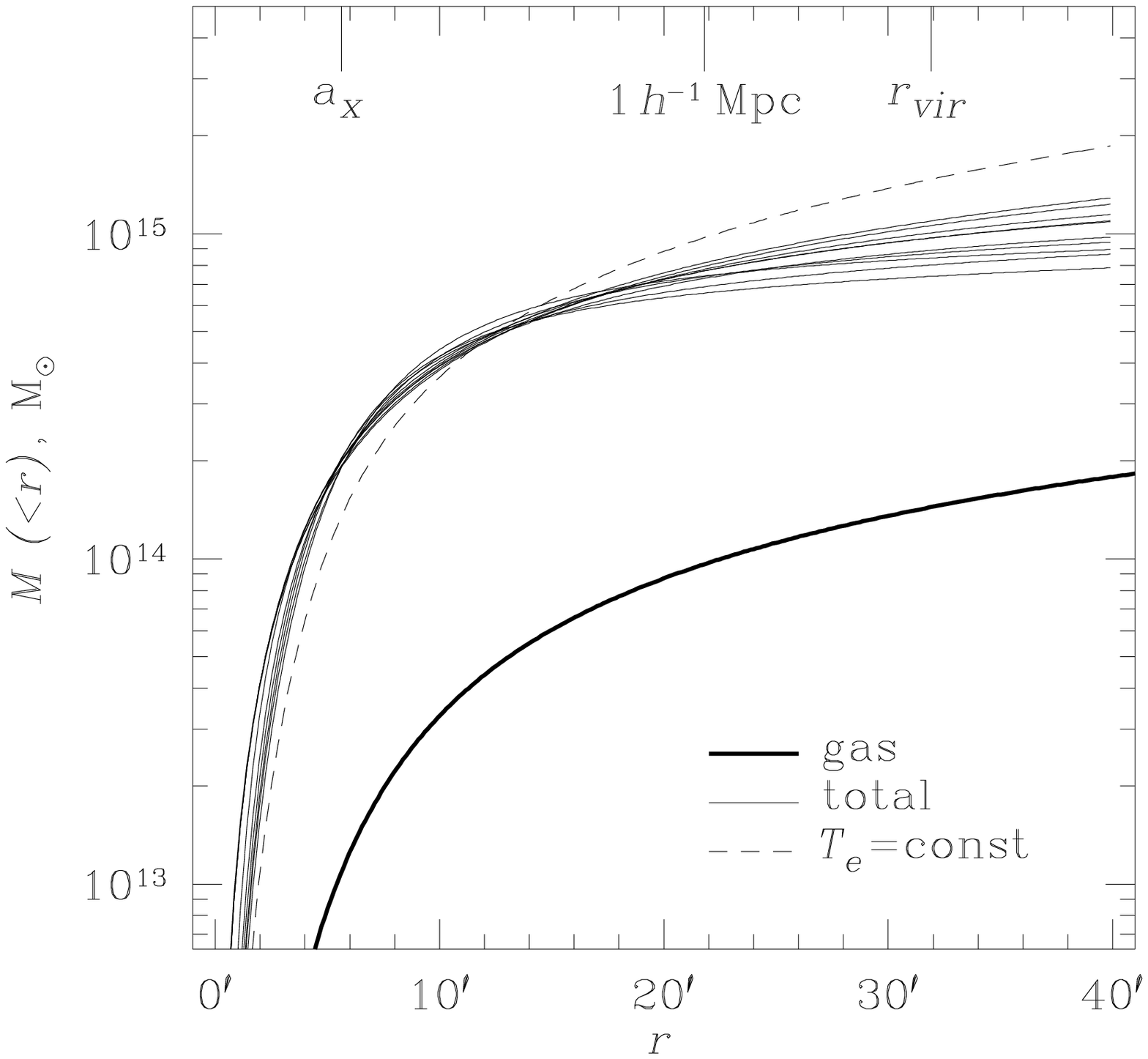}}

\rput[tl]{0}(0.15,11.5){\epsfysize=8.4cm
\epsffile{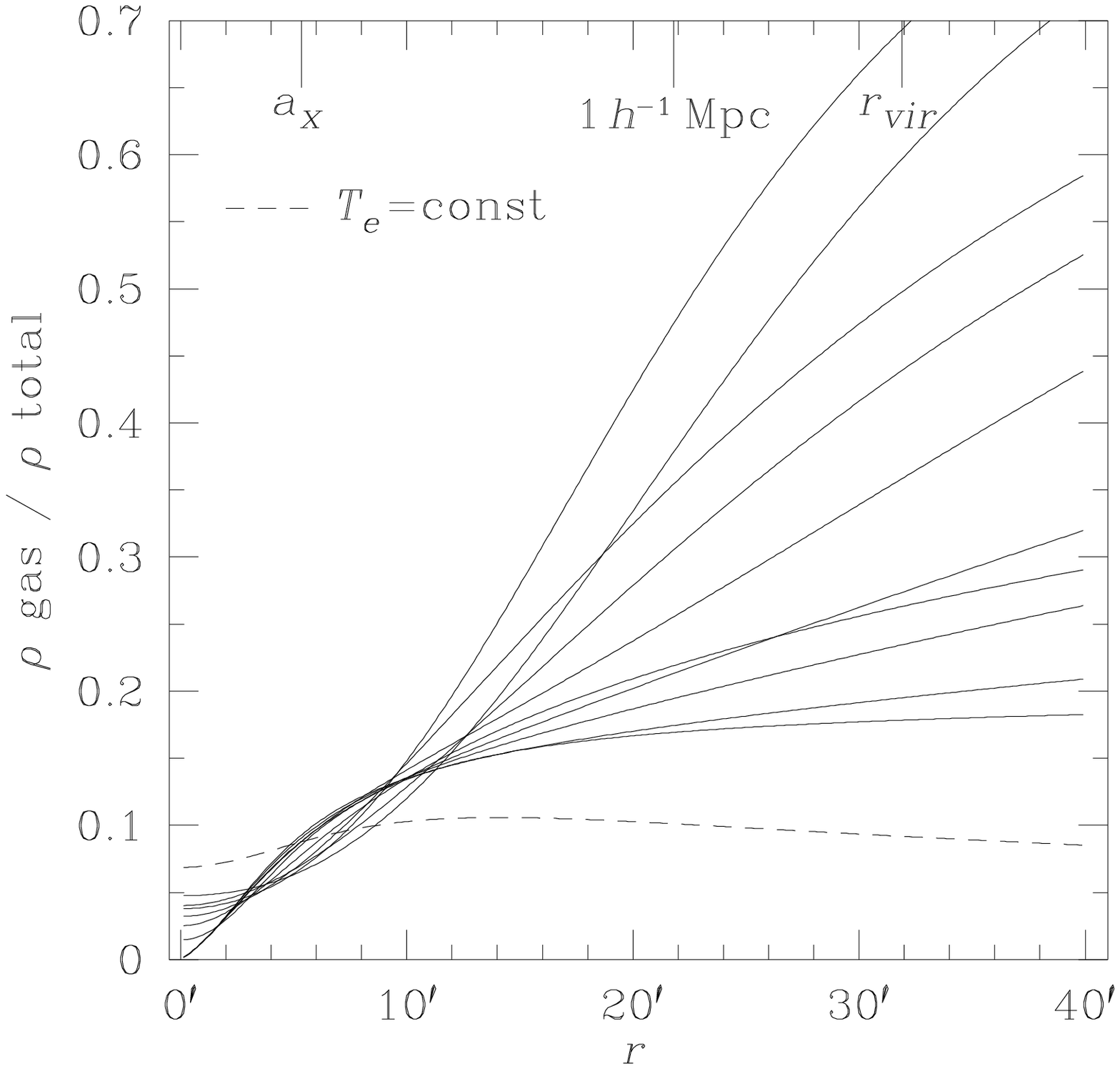}}

\rput[tl]{0}(9.25,11.5){\epsfysize=8.4cm
\epsffile{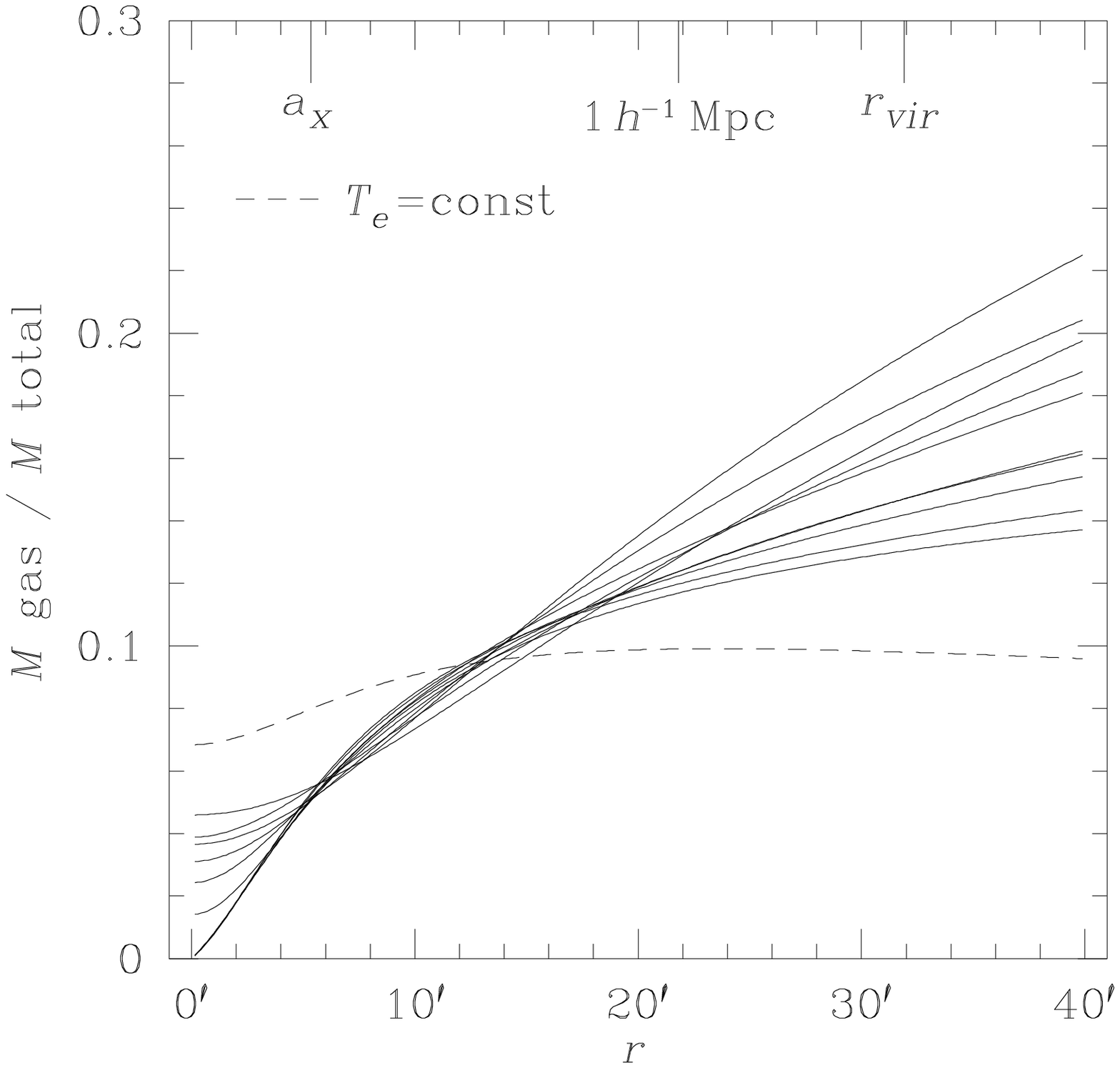}}

\rput[bl]{0}(8.3,19.25){\large\bf\em a}
\rput[bl]{0}(17.5,19.25){\large\bf\em b}
\rput[bl]{0}(8.3,10.65){\large\bf\em c}
\rput[bl]{0}(17.5,10.65){\large\bf\em d}

\rput[tl]{0}(-0.1,2.8){
\begin{minipage}{\textwidth}
\small\parindent=3.5mm
{\sc Fig.}~3.---Models whose temperature profiles are shown in Fig.\ 2:
({\em a}) their dark matter density profiles; ({\em b}) corresponding
enclosed total masses (including gas); ({\em c}) fractions of the gas
density in the total local density at a given radius; ({\em d}) fractions of
gas in the total mass within a given radius. Values from the isothermal
analysis (for the cluster average $T_e=7.5$ keV) are shown for reference.
Acceptable profiles of both functional forms (with a core or a central
$\rho_d \propto r^{-1}$ cusp) are shown for $\alpha\leq 5$. Absolute values
are for $h=0.65$ (total mass scales as $h^{-1}$ and gas fraction as
$h^{-3/2}$).
\end{minipage} 
}
\endpspicture
\end{figure*}

\subsection{Results}

A narrow range of hydrostatic temperature profiles is allowed by both
\asca\ and \rosat\ constraints. Representative examples of the temperature
profiles corresponding to the models acceptable within the approximate
combined 99\% confidence region as described above are shown in Fig.~2 (note
that the temperatures in the central $r\lax1'-2'$ may have little physical
meaning due to our incomplete treatment of the central galaxy atmosphere, as
discussed above). The corresponding dark matter density and total mass
profiles are shown in Fig.~3.  Shallow mass distributions best describe the
\asca\ temperatures alone, but they predict a temperature drop to zero
immediately beyond the \asca\ field, contrary to the \rosat\ data.  The
combined data exclude, at greater than 99\% confidence, dark matter profiles
with a core of the form (1) for which $\alpha<2.5$, and those of the form
(2) for which $\alpha<3$, for any value of the scale parameter $a_d$ (these
constraints on $\alpha$ in fact correspond to similar non-parametric slopes
of the profile over the measured region). Very steep dark matter profiles
with $\alpha>4$, which correspond to the outer cluster regions being
dominated by gas, are not excluded by the data, although they lack physical
motivation. Naturally, acceptable profiles with shallower outer slopes have
smaller $a_d$. For dark matter profiles with cores, $a_d/a_x\simeq 0.3-1$
(that is, dark matter is more centrally peaked than gas), and for profiles
with cusps, $a_d/a_x\simeq 1-2$ for the values of $\alpha\leq 5$ we
considered. For a particular value of $a_d= a_x\simeq r_{\rm vir}/6$, the
most acceptable outer slopes are $\alpha\simeq 4$ and 3 for the core and the
cusp profile forms, respectively; thus the ``universal'' profile of Navarro
et al.\ (1997) appears to be acceptable.

At $r=0.7\,h^{-1}$ Mpc (about half the virial radius) where direct accurate
temperature measurements exist, the total mass for the allowed models is
$3.8\pm 0.3 \times 10^{14}\,h^{-1}$\msun\ (for the mass values hereafter, we
include in quadrature the uncertainties of the local gas density slope at
the respective radii, as estimated in \S2). This mass is consistent with the
value of $4.5\pm 2 \times 10^{14}\,h^{-1}$\msun\ obtained earlier by HBN at
this radius (correcting for their distance error).

At the radius of the mean overdensity 500, which for A2256 corresponds to
$r=0.92\,h^{-1}$ Mpc using our mass profile, the mass is $4.5\pm 0.4 \times
10^{14}$\msun, significantly below the value of $7.2\pm 1.9\times
10^{14}$\msun\ predicted by the scaling law of Evrard et al.\ (1996). This
difference is due to the fact that the simulations typically produce a
shallower temperature fall than the observed one. If the cluster had been
isothermal, its mass within $r_{500}$ would be $6.7\times 10^{14}\,
h^{-1}$\msun, in agreement with the scaling relation (in part because
$r_{500}$ itself would be slightly larger).

At $r=1.5\,h^{-1}$ Mpc (the virial radius), we obtain an unprecedentedly
accurate estimate of the mass, $M=6.0\pm 1.5\times 10^{14}\,h^{-1}$\msun.
This value is significantly smaller, by a factor of 1.6, than the value
obtained assuming isothermal gas, as is naturally expected for a declining
temperature profile (e.g., Evrard et al.\ 1996). The local dark matter
density is constrained less accurately, but it is possible to derive a
strong upper limit beyond $r \sim 0.5\,h^{-1}$ Mpc. This limit is 1.5--2
times lower than the isothermal value, although it is of course rather
model-dependent.

Because the central cluster temperature is poorly constrained, so is the
density profile within the cluster X-ray core --- the profiles with and
without a central density cusp are both acceptable. Nevertheless, the total
mass internal to the core ($r=0.26\,h^{-1}$ Mpc) is well constrained as
$1.28 \pm 0.08\times 10^{14}\, h^{-1}$\msun\ for the acceptable models.
This is consistent with the {\em BBXRT\/} measurement (restricted to
$r<0.4\,h^{-1}$ Mpc; Miyaji et al.\ 1993).

\section{DISCUSSION}

\subsection{Accuracy of mass values}

The formal confidence intervals for the values given above represent only
the experimental uncertainties (and as such are directly comparable to those
from earlier works). Because of the high quality of the data, additional
uncertainties of the method due to possible deviations from spherical
symmetry and hydrostatic equilibrium, as predicted by simulations (\S4.2),
become comparable to or even exceed our experimental errors. We do not
attempt to quantify these systematic uncertainties for A2256, since the main
purpose of this paper is to compare the mass estimates obtained using the
actual temperature profile with those from the isothermal analysis (for
which the same systematic deviations apply). To draw reliable quantitative
conclusions about the cluster mass profiles in general (which is our more
distant goal), it is obviously necessary to study a sample of clusters, of
which A2256 is the first member. Nevertheless, even for this individual
cluster, the qualitative effects discussed below exceed the realistic
uncertainties suggested by the simulations.

It is also noteworthy that the formal measurement accuracy is approaching
the limit due to the unknown cluster peculiar velocities and hence accurate
distances, $d$. The two components of A2256 have velocities differing by
about 2000\kms and thus their distances calculated from redshifts may be in
error by as much as 5--7\%, implying a similar error in mass within a given
projected radius. In the absolute sense, this uncertainty is small and, of
course, $H_0$ is not yet known to such accuracy. However, relative results
may be significantly affected. For example, the observed difference of
baryon fractions (which are proportional to $d^{3/2}$) between the low-$z$
clusters A1060 and AWM7 (Loewenstein \& Mushotzky 1996a) might be entirely
explained if either of them had a peculiar velocity of 800--1000\kms.

\subsection{Gas fraction}

The mass value we obtain within $r_{\rm vir}$ corresponds to a factor of 1.6
higher average baryon fraction (which is higher or equal to the gas
fraction, $0.08\pm 0.02\,h^{-3/2}$) than the value obtained from the
isothermal analysis. This change, if applicable to other clusters, would
make the observed cluster baryon fraction even more contradictory with the
hypothesis of $\Omega=1$, as discussed e.g., by White et al.\ (1993), White
\& Fabian (1995), and most recently by Evrard (1997).  The gas fraction
above formally implies that $\Omega\simeq 0.1-0.25$ for $h=0.65$ if one
assumes that the cluster matter content is representative of the Universe as
a whole, that dark matter is nonbaryonic, and the standard nucleosynthesis
(Walker et al.\ 1991).  Alternatively, assuming the high or low measured
deuterium abundance (see review in Steigman 1996) it implies $\Omega\simeq
0.06-0.13$ and 0.2--0.5, respectively.
 
As Figures 3({\em c,d}) show, the gas fraction increases significantly with
radius, by about a factor of 2.5--3 from the radius of overdensity $10^4$
($r\sim a_x$) to $r_{\rm vir}$ even if one considers only the least steep
models. Similar behavior is obtained for a sample of other clusters under
the assumption of gas isothermality and for several cool systems taking into
account the observed temperature gradients (David, Jones, \& Forman 1995).
Note that A2256 differs from clusters in that sample by its steeper gas
density profile and a considerably lower {\em isothermal} estimate of the
gas fraction (which is also almost uniform with radius). However, the use of
the actual temperature profile makes the value and the radial dependence of
the gas fraction in A2256 rather similar to that for the rich clusters from
David et al.\ (if the actual temperature profiles would not alter their
results).  As David et al.\ point out, differences in the distributions of
gas and dark matter indicate that in addition to gravity, reheating and
hydrodynamics must play a role in the formation of structure on cluster
scales.  Loewenstein \& Mushotzky (1996a) similarly interpret the observed
difference (by a factor of $\sim 1.5$) of the individual baryon fractions in
the central parts of A1060 and AWM7. Cluster simulations including shock
heating of gas (e.g., Pearce, Thomas, \& Couchman 1994; Navarro et al.\
1995) plus radiative cooling (Frenk et al.\ 1996) but not including any
additional sources of gas heating, indeed produce slightly antibiased gas
profiles ($M_{\rm gas}/M_{\rm total} \propto r^{\,0.1-0.2}$, Evrard 1997) due
to the energy transfer from the dark matter and conversion of gas during
galaxy formation.  However, they do not produce the antibias of the observed
degree, suggesting that some physics is still missing from the simulations.
The most frequently mentioned candidate is energy input from supernovae (see
Sarazin 1988 for a review).  Along a different line, e.g., David, Forman, \&
Jones (1991) and Loewenstein \& Mushotzky (1996b) showed that the supernovae
needed to reproduce the elemental abundances in clusters would also produce
the amount of heat comparable to the gas thermal energy.  Still, the
published cluster simulations that attempt to model supernovae-driven
galactic winds (e.g., Metzler \& Evrard 1994) seem to fail to reproduce the
observed difference between the gas and dark matter distributions.

\subsection{Shape of the mass profile; lensing masses}

At present, our constraints on the mass profile can be directly compared
only to another \asca\ result, that on A2163 (M96a), since other work was
limited by the unavailability of the temperature data at large radii.  The
constraints on A2163 were rather similar to those obtained here (total
density profiles of the form (1) with $\alpha<2.1$ were excluded,
corresponding to a slightly stronger constraint on the dark matter slope),
although they were less accurate due to A2163's greater distance and less
conclusive due to the merger in that system. Using an independent method
based on the analysis of ellipticity of cluster X-ray images, Canizares \&
Buote (1997) infer that the density of dark matter in clusters falls off as
$\rho_d \propto r^{-4}$ in the outer parts. This is very similar to what we
find for A2256.

The shape of the dark matter halo may be an interesting cosmological probe
(e.g., Hoffman \& Shaham 1985; Crone et al.\ 1994; see, however, Navarro et
al.\ 1997). Comparing a sample of just one cluster with the statistical
conclusions of the simulations is uncertain and has to wait until more
clusters are studied. It is interesting to note in the meantime that the
shallow density profiles predicted by Kofman et al.\ (1996) for the CHDM
universe are inconsistent with our A2256 results.

Gravitational lensing provides an independent measure of the cluster mass
distribution (see review in Bartelmann \& Narayan 1995). Often, masses
inferred by this method are up to a factor of a few greater than those from
X-rays in the central cluster regions (e.g., Loeb \& Mao 1994; Wu 1994;
Miralda-Escud\'e \& Babul 1995; Tyson \& Fischer 1995). In some (the most
discrepant) cases, the lensing analysis is likely to overestimate the mass
as a result of substructure (Bartelmann 1995) or projection (Daines et al.\
1997). Bartelmann \& Steinmetz (1996) showed that clusters selected for the
presence of strong arcs are more likely to have substructure. However, with
better temperature data, the X-ray mass estimates may also be revised
upwards in better agreement with lensing measurements, as we find for A2256.
Our inferred density profiles for both functional forms are more centrally
peaked than the gas profile, in agreement with lensing (Bartelmann \&
Narayan 1995) and the A2163 finding (M96a). Note that profiles with high
central densities would erroneously be {\em excluded} in our analysis if one
did not allow for the multicomponent spectrum in the central region
(required by \asca) and instead used a formally well-constrained single
temperature that artificially excludes the possibility of a hotter gas. This
is similar to the result of Allen et al.\ (1995) who find that allowing for
a multi-temperature gas in the cooling flow region of PKS0745 increases the
X-ray-measured mass within the lensing radius by a large factor, such that
the strong lensing and X-ray masses agree. There are no lensing data on
A2256 to compare because it is too nearby. Ikebe et al.\ (1996, 1997)
similarly note that central dark matter concentrations in the cD clusters
Fornax and Hydra-A are required if the gas in their central regions is
treated as multi-phase as required by the \asca\ spectra.

The fact that the shallow dark matter distributions are excluded at large
radii may also have implications for the weak lensing mass estimates. These
estimates at present are of a relative nature and require assumed model
density profiles in order to obtain the absolute value of the mass in the
central region. As Squires et al.\ (1996) note, the central mass inferred
from their analysis of A2218 varies by a factor of 1.3 depending on the
assumed density profile, with the steeper profiles resulting in lower
values.  Our results on A2256 prefer steep density profiles, hence favor the
lower bound of the weak lensing mass estimates.

Regarding the dark matter distribution near the center, we note that in a
cluster whose gas has a constant density core and is supported only by its
own pressure, a central dark matter cusp may be ruled out a priori because
it would require a convectively unstable central temperature peak. A2256 has
a central density excess over the $\beta$-model, associated with the
brightest galaxy, which we do not model adequately (for a given mass
profile, it would reduce the hydrostatic central temperature), which is why
this additional constraint was not applied here. For other clusters, this
constraint can be used to devise an interesting test of the significance of
nonthermal gas support (which was proposed as one explanation of the
discrepant lensing and X-ray masses; Loeb \& Mao 1994). If a relaxed cluster
is found whose gas indeed follows a $\beta$-model all the way to the center
but whose lensing data nevertheless require a central density cusp, then the
gas in the center must have nonthermal support.

\section{SUMMARY}

We constrained the mass distribution in the primary cluster of A2256, using
actual gas temperature measurements at large radii derived from the X-ray
data of unsurpassed quality for any cluster to date. Our analysis has shown
that:

1. Dark matter density profiles steeper than $\rho_d \propto r^{-2.5}$ in
the outer part are required by the data. Profiles with central cusps such as
those predicted by simulations and implied by lensing results are consistent
with the data. The mass is distributed rather differently from what may be
inferred under the assumption of isothermal gas: the total mass internal to
the cluster X-ray core is higher, while the mass within the virial radius is
lower, both by factors of about 1.5, than the isothermal estimates;

2. If our results for A2256 are typical for other clusters, three
particularly interesting implications would be that the X-ray and lensing
masses converge; the ``baryon catastrophe'' is even more pronounced; and
gravity and merger shocks are probably not the only significant source of
heating of the cluster gas.

\acknowledgments

We thank \asca\ and \rosat\ teams for their work on obtaining the data we
used here. We are grateful to W. Forman, R. Mushotzky and the referees, one
of whom was M. Henriksen, for valuable comments on the manuscript. This work
was supported by NASA grant NAG5-2611 and by CfA postdoctoral fellowship.

\appendix
\section{SOLUTION OF THE HYDROSTATIC EQUILIBRIUM EQUATION}

The hydrostatic equilibrium equation for a spherically symmetric cluster
(e.g., Sarazin 1988),
\begin{equation}
\int^r_0 4\pi r^2 (\rho_d+\rho_g)\; dr\; = \; 
3.70\times 10^{13} M_\odot\, \frac{T(r)}{\rm 1\;keV}\,
\frac{r}{\rm 1\;Mpc}\, \frac{0.60}{\mu}
\left(- \frac{d \ln \rho_g}{d \ln r} - \frac{d \ln T}{d \ln r} \right),
\end{equation}
can be rewritten as a differential equation for the gas temperature,
assuming the gas density profile of the form (3):
\begin{equation}
\frac{dt}{dx}=\frac{3\beta x}{1+x^2}\,t-\frac{A}{T_0}\frac{I(x)}{x^2}.
\label{diff}
\end{equation}
Here we denote $x\equiv r/a_x$; $t\equiv T/T_0$ where $T_0$ is the gas
temperature in keV at $r=0$, so that $t(0)\equiv 1$;
\begin{equation}
A\equiv \frac{4\pi}{3.70\times 10^{13}} \left(\frac{a_x}{\rm 1\;Mpc}\right)^2 
\frac{\rho_{g0}}{1\;M_\odot {\rm Mpc}^{-3}}\; {\rm keV};
\end{equation}
and
\begin{equation}
I(x)\equiv \int^x_0 y^2 (1+y^2)^{-3\beta/2}\;dy\; + \; 
\frac{\rho_{d1}}{\rho_{g0}}\; \int^x_0 y^2 f_d(y)\; dy.
\label{int}
\end{equation}
The two integrals in the equation above correspond to the gas and the dark
matter, respectively; $f_d\equiv \rho_d/\rho_{d1}$ (where $\rho_{d1}$ is the
dark matter density at the X-ray core radius $a_x$) denotes the
dimensionless dark matter density whose form is given by either (1) or (2).
The integrals in (\ref{int}) can be worked out analytically for the density
profiles we use, but it is faster to calculate them numerically directly
than evaluate numerically the resulting special functions.

Equation (\ref{diff}) can be solved either numerically (as e.g., in Hughes
1989) or analytically, as we do. The analytic solution is given by
\begin{equation}
t(x)=(1+x^2)^{3\beta/2}\left[
1-\frac{A}{T_0} \int^x_0 (1+y^2)^{-3\beta/2}\, \frac{I(y)}{y^2}\, dy \right].
\label{analyt}
\end{equation}
For a set of parameters $a_d$, $\alpha$, $\eta$, $\rho_{d1}$ and $T_0$, we
calculate the temperature profile using eq.\ (\ref{analyt}).  The parameters
$\rho_{d1}$ and $T_0$ are adjusted so that the profile approximates the
\asca\ data as described in \S5. In practice, it is advantageous to fit the
quantities $\rho_{d1}/T_0$ and $T_0$ rather than the above two, since in the
limit of zero gas mass the shape of the profile depends only on
$\rho_{d1}/T_0$.

The scheme described here differs from the one used by Hughes (1989), HBN
and M96a in that we parameterize a dark matter profile rather than a total
mass profile. This automatically excludes the unphysical solutions in which
the dark matter density ($\rho_{\rm total}-\rho_g$) is negative, which
required the introduction of an additional $H_0$-dependent constraint.

\end{document}